 \documentclass[final,1p,times]{elsarticle}
\usepackage{amssymb}
\newtheorem{theorem}{Theorem}

\def \Degree{\mathop{\rm deg}\nolimits}
\def \Order{\mathop{\rm order}\nolimits}

\def \ccomma{\raise 2pt\hbox{,}} % Le petit livre de TeX page 234
\def \D {\hbox{d}}

\def \sn {\mathop{\rm sn}\nolimits}
\def \cn {\mathop{\rm cn}\nolimits}
\def \dn {\mathop{\rm dn}\nolimits}
\def \sech{\mathop{\rm sech}\nolimits}

\def \Im  {\mathop{\rm Im}\nolimits}
\def \mod#1{\vert #1 \vert}

\def \GLA{A}
\def \GLAc{\overline{\GLA}} % not nice
\def \GLAc{\bar\GLA}
\def \GLa{a}

\def \poleorder{s} % p would be DOUBLY DEFINED

\def \csi{\kappa_{\rm i}} 
\def \csr{\kappa_{\rm r}} 
\def \tauone{\tau_1}
\def \taub{\tau_b}

\def \xib  {\xi_b}
\def \Kshift{\kappa} % CGL5 front = K - \csr/2
 
\def \Cthree{K_1} 
\def \Cfour {K_2} 
\def \Mshift{M_0} % very bad notation for $a_2 k_0$
\def \ersurei{\lambda}            
 
\def \ex{e_1} % ( c s_i)^2/48.                     
\def \ey{e_0} %  g_r/36.                           

\def \jmax{J}

\journal{Physics letters A}

\begin{document}

\begin{frontmatter}


\title{All meromorphic traveling waves of cubic and quintic complex Ginzburg-Landau equations}
%\title{The eleven meromorphic traveling waves of cubic and quintic complex Ginzburg-Landau equations}
%
%use optional labels to link authors explicitly to addresses:
% ERROR of elsarticle.cls if blanks like in \author[n ]
\author[Borelli,HKUMath]{Robert Conte\corref{cor1}}\ead{Robert.Conte@cea.fr,         ORCID https://orcid.org/0000-0002-1840-5095}
\author[VUBTENA]        {Micheline Musette}        \ead{Micheline.Musette@gmail.com, ORCID https://orcid.org/0000-0002-2442-9579}
\author[HKUMath]        {Tuen Wai Ng}              \ead{NTW@maths.hku.hk,            ORCID https://orcid.org/0000-0002-3985-5132}
\author[SZUIAS,SZUCMS]  {Chengfa Wu}               \ead{CFWu@szu.edu.cn,             ORCID https://orcid.org/0000-0003-1697-4654}
\cortext[cor1]{Corresponding author.
Robert Conte,
Centre Borelli,
\'Ecole normale sup\'erieure Paris-Saclay,
4 avenue des sciences,
F-91190 Gif-sur-Yvette,
France.
Robert.Conte@cea.fr
}
%
%affiliation[Borelli]{organization={Universit\'e Paris-Saclay, ENS Paris-Saclay, CNRS}, \'e => 6 errors of elsarticle.cls
\affiliation[Borelli]{organization={Universite   Paris-Saclay, ENS Paris-Saclay, CNRS},
            addressline={Centre Borelli},
            city={Gif-sur-Yvette},
            postcode={91140},
           %state={},
            country={France}}
\affiliation[VUBTENA]{organization={Vrije Universiteit Brussel},
            addressline={Dienst Theoretische Natuurkunde, Pleinlaan 2},
            city={Brussels},
            postcode={1050},
           %state={},
            country={Belgium}}
\affiliation[HKUMath]{organization={The University of Hong Kong},
            addressline={Department of Mathematics, Pokfulam},
           %city={Hong Kong},
           %postcode={},
           %state={},
            country={Hong Kong}}
\affiliation[SZUIAS]{organization={Shenzhen University},
            addressline={Institute for Advanced Study},
            city={Shenzhen},
           %postcode={},
           %state={},
            country={PR China}}
\affiliation[SZUCMS]{organization={Shenzhen University},
            addressline={College of Mathematics and Statistics},
            city={Shenzhen},
            postcode={518060},
           %state={},
            country={PR China}}
\begin{abstract}
For both cubic and quintic nonlinearities of the one-dimensional
complex Ginzburg-Landau evolution equation,
we prove by a theorem of Eremenko the finiteness of the number of 
traveling waves 
whose squared modulus has only poles in the complex plane,
and we provide all their closed form expressions.
Among these eleven solutions, five are provided by the method used.
This allows us to complete the list of solutions previously obtained by other authors.
\end{abstract}



\begin{keyword}
complex cubic and quintic Ginzburg-Landau equation,
closed-form solutions,
Nevanlinna theory,
nonlinear optics,
turbulence,
traveling waves,
patterns,
coherent structures,
defect-mediated turbulence,
dark solitons.


\PACS
 02.30.Hq, % ordinary differential equations 
%02.30.Jr, % Differential equations, partial
 02.30.+g  % Mathematical methods in physics
% (cosmology, general relativity, nonlinear ODEs, quantum gravity)
            %   Function theory, analysis


\MSC %AMS MSC 2000 Mathematics Subject Classification: 
     % https://mathscinet.ams.org/msc/pdfs/classifications2020.pdf
%33E17, % Painlev\'e-type functions
%34Mxx, % Differential equations in the complex domain [See also 30Dxx, 32G34] 
 34M04, % Nonlinear ordinary differential equations and systems in the complex domai
%35A20, % PDEs. Analytic methods, 
 35Q99  % Equations of mathematical physics, none of the above

\end{keyword}

\end{frontmatter}


\bigskip
% https://www.elsevier.com/journals/physics-letters-a/0375-9601/guide-for-authors

An error in the style file elsarticle.cls prevents us to put the acute accent on Universit\'e.

\vfill\eject
\tableofcontents

% ------------------------------------------------------------------------
\section{Introduction}

In 1950 Ginzburg and Landau \cite{GL1950} introduced a description of superconductivity,
in the absence of an external magnetic field,
as a second order phase transition 
in which the order parameter is a complex function $\GLA(x,t)$
(which they denoted $\Psi$ for its connection with quantum mechanics)
deriving from a free energy quartic in $\mod{\GLA}$ (``$\varphi^4$ theory'').
This assumption naturally led them, in the one-dimensional case,
to an evolution equation invariant under a translation of the phase $\arg{\GLA}$,
now known as the one-dimensional cubic complex Ginzburg-Landau equation CGL3
\begin{eqnarray}
& &  {\hskip -18.0 truemm}
\hbox{(CGL3) }
i \GLA_t +p \GLA_{xx} +q \mod{\GLA}^2 \GLA -i \gamma \GLA =0,
%(\GLA,p,q,r) \in \mathbb{C},
p  \gamma \Im(q/p)\not=0,
%\gamma \in \mathbb{R},
\label{eqCGL3}
\end{eqnarray}
in which % $\Delta$ is the Laplacian operator,
$p,q$ are complex constants and $\gamma$ a real constant.

This CGL3 equation later turned out to be a generic equation 
arising from the approximation of a slowly varying amplitude,
with applications to quite various physical phenomena, such as 
spatio-temporal turbulence, 
Bose-Einstein condensation,
and more recently numerous fields of nonlinear optics,
as detailed in several reviews
\cite{MannevilleBook} % 1990
\cite{AK2002} %\showCGL{CGL3/5} I.S.~Aranson and L.~Kramer,  
\cite{vS2003} %\showCGL{CGL3/5} W.~van Saarloos,
\cite{Moloney-Newell-Book} % 2004
\cite{AABook2008} 
\cite{FerreiraBook2022}. 

CGL3 describes for instance the formation of patterns near a Hopf bifurcation,
$\gamma$ measuring the difference between the order parameter and its critical value.
When the bifurcation is subcritical,
the cubic term is insufficient to describe the system and
one must take account of the next nonlinearity compatible with
the phase invariance,
thus defining the complex quintic equation CGL5,
\begin{eqnarray}
& &  {\hskip -18.0 truemm}
\hbox{(CGL5) }
i \GLA_t +p \GLA_{xx} +q \mod{\GLA}^2 \GLA +r \mod{\GLA}^4 \GLA -i \gamma \GLA =0,
%(\GLA,p,q,r) \in \mathbb{C},
p r \gamma \Im(r/p)\not=0,
%\gamma \in \mathbb{R},
\label{eqCGL5}
\end{eqnarray}
in which $r$ is a complex constant.

The phase diagrams of both
CGL3 and CGL5 are quite rich
\cite[Fig.~1]{Chate1994} 
\cite[Fig.~1a]{vanHecke}
and comprise a variety of chaotic and regular phases.
Moreover, a remarkable feature is the existence,
observed in both computer and real experiments,
of a very small number of elementary patterns
able to describe most r\'egimes and, more importantly,
to act as separators between the different r\'egimes.
When they are traveling waves,
($c$ and $\omega$ real constants, $M$ and $\varphi$ real functions, $a$ complex function),
\begin{eqnarray}
& & {\hskip -15.0 truemm}
\GLA =\sqrt{M(\xi)} e^{ i(\displaystyle{-\omega t + \varphi(\xi)})}
=a(\xi) e^{\displaystyle{-i \omega t}},
\xi=x-ct,
\label{eqCGL35ReducMphi}
%\label{eqCGL35SystemMphi}
\end{eqnarray}
these coherent structures have been classified
\cite[Fig.~1]{vSH1992} according to 
their 
topology
(pulses, fronts, shocks, holes, sinks, defects, etc)
and the nature of their orbits:
homoclinic 
(equal values of $\lim_{x \to - \infty} \mod{A}$ and  
$\lim_{x \to + \infty} \mod{A}$) 
or heteroclinic (unequal values).

For instance, a CGL3 heteroclinic hole has been analytically found
by Bekki and Nozaki \cite{BN1985},
and the CGL3 homoclinic hole has only been observed in numerical experiments
by van Hecke \cite{vanHecke}
but not found analytically.

We restrict here to the situation
in which the ratio of the highest nonlinearity coefficient $r$ or $q$
by the dispersion coefficient $p$ is not real,
i.e.~when the CGL equation is dissipative \cite{AABook2008}.
Indeed,
when this ratio is real, the behaviour is no more dissipative
but dispersive and, at least in the cubic case, the CGL equation then has 
the same singularity structure 
than the nonlinear Schr\"odinger equation (NLS).
While the exact solutions of NLS are numerous,
very few exact solutions are known in the dissipative case.

Before the introduction of the method described in section \ref{sectionNecessary},
only six analytic expressions for traveling waves were known:
a heteroclinic front,
a homoclinic pulse and
a heteroclinic source, 
for both CGL3 and CGL5,
they are recalled in the appendix.

In the present paper, we enumerate \textit{all} those traveling waves
which belong to a rather natural class.
Indeed, the six just mentioned exact traveling waves
share a nice property:
the only singularities of their squared modulus $M$
which depend on the initial conditions (in short, \textit{movable} singularities)
are poles, in the complex plane of course.
Conversely, if one requires the movable singularities of $M$ to be only poles
(in short, $M$ to be meromorphic on $\mathbb{C}$),
there exists a mathematical method able to yield all the resulting values of $M$
in closed form,
and our main result is the following.

\begin{theorem}
The CGL3 and CGL5 equations admit exactly eleven different traveling wave solutions
in which $M$ is meromorphic on $\mathbb{C}$,
and their closed form is known.
\label{Theorem-All-mero}
\end{theorem}
Table \ref{TableMeroCGL} 
(the vocabulary of its legend will be defined below)
displays the main features of these traveling waves,
in which the real parameters $d_r$, $d_i$, $e_r$, $e_i$, $\csr$, $\csi$, $g_r$ ,$g_i$,
equivalent to $p,q,r,\gamma,c,\omega$, are defined by
\begin{eqnarray}
\begin{array}{ll}
\displaystyle{
\frac{q}{p}=d_r + i d_i, 
\frac{r}{p}=e_r + i e_i,
\frac{c}{p}=\csr - i \csi,
%}\\ \displaystyle{
\frac{\gamma + i \omega}{p}=g_r + i g_i - \frac{1}{2} \csr \csi - \frac{i}{4} \csr^2.
}
\end{array}
\label{eqCGL35Notation}
\end{eqnarray}

%tabcolsep=1.0truemm
\tabcolsep=0.5truemm

\begin{table}[ht] % [p]
\caption[The 11 meromorphic solutions $M$ of CGL5 and CGL3.]{
         The 11 meromorphic solutions $M$ of CGL5 and CGL3,
				ordered by
				type (elliptic, trigonometric, rational)
				and number of poles of $M$.
				Columns display:
				codimension 
			%(number of real constraints on $p,q,r,\gamma$),
				(number of real constraints on $\csi, g_r, g_i, d_r, d_i, e_r$ (CGL5)   % $e_i,\csr$ excluded
				                            or $\csi, g_r, g_i,           d_r$ (CGL3)), % $d_i,\csr$ excluded
			%arbitrary movable parameters ($c$ or $\omega$),
			% number of Laurent series (of $M$ and $(\log \GLa)'$) involved,
				number of poles ($n$P$p$ means $n$ poles of order $p$) 
				in the Hermite decomposition of $M$ and $(\log \GLa)'$,
				solution $\GLA$ and its number of branches, topology, reference.
}
\vspace{0.2truecm}
\begin{center}
\begin{tabular}{| c | l | c | c | c | c | c | l | r |}
\hline % \hline % ********************************************************
CGL&Type & Codim&H($M$)&H($(\log\GLa)'$)& Sol       & Branches & Topology & Ref
\\ \hline \hline % ********************************************************
3 & Ellip& 2    & 2P2  & 4P1 & \cite[(3.181)]{CMBook2}    & 1 & Unbounded, periodic & \cite{CMBook2} \\ \hline \hline % *****
3 & Trigo& 1    & 1P2  & 2P1 & (\ref{eqCGL3HoleA})${}_1$  & 2 & Heteroclinic source/hole & \cite{BN1985} \\ \hline %\hline % ***** 
3 & Trigo& 2    & 1P2  & 1P1 & (\ref{eqCGL3PulseA})${}_2$ & 2 & Homoclinic pulse         & \cite{HS1972} \\ \hline %\hline % *****
3 & Trigo& 2    & 1P2  & 1P1 & (\ref{eqCGL3FrontA})${}_3$ & 2 & Heteroclinic front       & \cite{NB1984} \\ \hline \hline % *****
5 & Ellip& 4    & 4P1  & 3P1 & (\ref{eqCGL5Ellip-dloga})  & 1 & Unbounded, periodic      & \cite{ConteNgCGL5_ACAP} \\ \hline \hline % ** 
5 & Trigo& 2    & 1P1  & 1P1 & (\ref{eqCGL5FrontA})${}_1$ & 4 & Heteroclinic front       & \cite{vSH1992} \\ \hline %\hline % ********
5 & Trigo& 3    & 2P1  & 3P1 & (\ref{eqCGL5SourceA})${}_2$& 2 & Heteroclinic source/sink & \cite{MCC1994} \\ \hline %\hline % *
5 & Trigo& 3    & 2P1  & 2P1 & (\ref{eqCGL5PulseA})${}_3$ & 2 & Homoclinic pulse         & \cite{vSH1992} \\ \hline %\hline % **
5 & Trigo& 5    & 4P1  & 5P1 & \cite[(11)]{CMNW-CGL35-Letter}& 1 & Homoclinic defect& \cite{CMNW-CGL35-Letter} \\ \hline %\hline % ** 
5 & Trigo& 5    & 4P1  & 6P1 & (\ref{eqCGL5BoundA})       & 2 & Homoclinic bound state   & \cite{CMNW-CGL35-Letter} \\ \hline \hline % ** 
5 & Rat.l& 5    & 4P1  & 6P1 & (\ref{eqCGL5RatA})         & 2 & Unbounded                & Here               \\ \hline \hline % 
%
% ********************************************************
\end{tabular}
\end{center}
\label{TableMeroCGL}
\end{table}
	
This provides us with five more (and only five) 
analytic traveling wave solutions.
Three of them are unbounded and therefore unphysical,
and the two others \cite{CMNW-CGL35-Letter} represent coherent structures
already observed.

The first one is a CGL5 topological defect.
The occurence of defects 
\cite{CGL1989} % CGL3 only
\cite{vHH2001} % CGL3 only
is a major mechanism 
\cite{Shraiman-et-al1992} % CGL3 only
of transition to a turbulent state.
Although this ``defect-mediated turbulence'' 
has been mostly documented in two-dimensional CGL3 
\cite{LegaThese} \cite{Lega2001}, 
% 2001 \cite{Lega2001} page 271 \P 3: 2-dim CGL3 displays defect-mediated turbulence.
there exists a range of parameters of CGL5,
which includes the values of the exact defect,
displaying a similar ``hole-mediated turbulence''
\cite[Fig.~3b]{PSAWK1993} \cite[Fig.~4]{PSAK1995}
\cite[p 278]{Lega2001}:
for a destabilizing CGL5 term 
(negative $\delta$ in the notation of Ref.~\cite{PSAWK1993})
one observes a succession of phase slips (every time $M$ vanishes),
which create hole-shock collisions,
ending in a process of as many annihilations as creations.

The second bounded traveling wave is a bound state of two CGL5 dark solitons,
which has been observed in numerical simulations \cite[Fig.~4]{ACM1998}.

The paper is organized as follows.
Section \ref{sectionSufficient} reviews the methods which succeeded to find some solutions,
and also unsuccessful methods, with the reasons for their failure.
Section \ref{sectionSingularities} recalls all the Laurent series of $M$,
whose knowledge is a prequisite for what follows.
The method to find all meromorphic solutions is presented in section \ref{sectionNecessary}.
Next section \ref{sectionNecessaryCGL} enumerates the five such CGL solutions
which were found by this method.
%with full details in one elliptic case (CGL3)
%and one trigonometric case (CGL5 defect).

% ------------------------------------------------------------------------
\section{Previous methods}
\label{sectionSufficient}

Quite different methods have been used to try and obtain closed form 
traveling wave solutions of CGL3/5.
The succesful ones are the following.

\begin{enumerate}
	\item 
The assumption $\GLA$ equal to the product of a function of $t$ by a real function of $x$
immediately yields the homoclinic pulse (\ref{eqCGL3Pulse})  
of CGL3 \cite{HS1972}.

	\item 
The Hirota method \cite{Hirota1980} consists in writing, when this is possible, the original 
partial differential equation (PDE)
only with the so-called Hirota derivation operators $D_x$ and $D_t$,
i.e.~without the usual derivation operators $\partial_x$, $\partial_t$.
Such a writing, which is indeed possible for CGL3 \cite{NB1984},
implies \textit{ipso facto} the existence of various solitary waves.
This allowed Bekki and Nozaki to obtain
a CGL3 front (\ref{eqCGL3Front}) \cite{NB1984}
and
a CGL3 hole (\ref{eqCGL3Hole}) with an arbitrary velocity \cite{BN1985}.
	
	\item 
The replacement of the 	
third order ODE (\ref{eqCGL35Order3}) for $M(\xi)$
by an equivalent polynomial dynamical system in three real components $(M,L=M'/M,\psi=\varphi'-\csr/2)$
\cite{vSH1992}
\begin{eqnarray}
& &
\frac{\D}{\D \xi} \pmatrix{M \cr L \cr \psi \cr}=
\pmatrix{
M L \cr 
-\frac{3}{4} L^2 + \csi L + 2 \psi^2-2 e_r M^2-2 d_r M -2 g_i \cr 
- L \psi + \csi \psi - e_i M^2 - d_i M + g_r \cr
},
\label{eqCGL5ReducRealDynamicalSystem}
\end{eqnarray}
followed by heuristic constraints on these three components.
This allowed van Saarloos and Hohenberg to discover the CGL5 front (\ref{eqCGL5Front})
and the CGL5 pulse (\ref{eqCGL5Pulse}),
with the respective constraints \cite[Eqs.~(3.38), (3.51)]{vSH1992}
($c_j,d_j$ adjustable constants),
\begin{eqnarray}
% (2.5)-(2.6)=3-dim poly system for (a=M,q=phi',kappa=a'/a)
% (3.38)=assumption q and kappa linear in a^2 => CGL5 front (new)
% (3.51)=assumption kappa^2=1+a^2+a^4, q=1+kappa  => CGL5 pulse (new)
& & \hbox{(front) } 
\psi=c_1+c_2 \frac{M'}{M}, \frac{M'}{M}=c_3+c_4 M,
\ '=\frac{\D}{\D \xi}\ccomma
\\ 
& & \hbox{(pulse) }  
\psi=d_1+d_2 \frac{M'}{M}, \frac{{M'}^2}{M^2}=d_3+d_4 M+d_5 M^2.
\end{eqnarray}

	\item 
Truncation methods as initiated by Weiss \textit{et al.} \cite{WTC}.
By implementing an extension \cite{MC1994} of the WTC truncation method,
Marcq \textit{et al.} \cite{MCC1994} found the CGL5 source or sink (\ref{eqCGL5Source}).

	\item 
The enforcement by Hone \cite{Hone2005}, for nondegenerate elliptic solutions,
of the classical ne\-cessary condition that the sum of the residues of one or more Laurent series % (\ref{eqCGL3-M-series})
of $M$ 
(or more generally of any rational function of derivatives of $M$)
inside a period parallelogram vanishes.
This method allowed Vernov \cite{VernovCGL5}
to find an elliptic solution of CGL5,
however not the most general one for reasons explained in Section \ref{sectionCGL5Elliptic}.

\end{enumerate}

	Since tremendous efforts 
\cite{Jones-etal1989} % [CGL5] Existence of CGL5 fronts.
\cite{GB1991} % [CGL5] L.~Gagnon and P.A.~B\'elanger,
\cite{Lega2001} % CGL3
have been made to search for additional closed form solutions,
it is also worth mentioning why other methods failed in the case of CGL.

\begin{enumerate}

	\item 
	The search for elliptic solutions can be made by assuming the squared modulus
to be a
polynomial of $\sn, \cn,\dn$, 
or a polynomial of $\wp,\wp'$ \cite{KudryashovElliptic}. % 1989
% Samsonov, preprints 1259 and 1293 (elliptic solutions to PDEs), Ioffe, Leningrad (1988).
%  never published as said by Samsonov in https://doi.org/10.1134/S1063771010060114 (2010)
Since elliptic solutions are generically not polynomials of such functions,
this explains the failure for CGL of this too restrictive assumption.
	
	\item 	
	Similarly, another search for elliptic solutions \cite{Klyachkin1989} 
by requiring the squared modulus $M$ to obey the most general first order second degree
binomial ODE of Briot and Bouquet 
${M'}^2=\sum_{j=0}^4 c_j M^j$ fails to find any elliptic solution
and only finds various known trigonometric degeneracies.
Indeed, this assumption can only yield homographic functions of
$\wp,\wp^2,\wp^3,\wp'$.

	\item 
	Innumerable ``new methods'' are regularly proposed, % published
such as the 
``Exp-method'', 
``$G'/G$ expansion method'', 
``simplest equation method'', 
``homogeneous balance method'', 
etc,
but they are just copies of the previously mentioned methods,
see the criticisms in Refs
\cite{Be-careful} 
and
\cite{More-common-errors}. 

\end{enumerate}

% ------------------------------------------------------------------------
\section{Movable singularities of CGL}
\label{sectionSingularities}

Traveling waves (\ref{eqCGL35ReducMphi})
are characterized by a third order nonlinear
ordinary differential equation (ODE) for the squared modulus $M(\xi)$ 
\cite[p.~18]{Klyachkin1989} \cite{MC2003},
\begin{eqnarray}
& & %{\hskip -15.0 truemm}
%\left\lbrace
\begin{array}{ll}
\displaystyle{
(G'-2 \csi G)^2 - 4 G M^2  (e_i M^2 + d_i M - g_r)^2=0,
}\\ \displaystyle{
G=\frac{M M''}{2} - \frac{M'^2}{4} 
  -\frac{\csi}{2} M M'  + g_i M^2 + d_r M^3 + e_r M^4, 
}\\ \displaystyle{
\varphi' =\frac{\csr}{2}+\frac{G'-2 \csi G}{2 M^2( g_r - d_i M - e_i M^2)}\cdot
%\varphi' =\frac{\csr}{2}+\psi,
%\psi  =\frac{G'-2 c s_i G}{2 M^2( g_r - d_i M - e_i M^2)},
%\psi^2=\frac{G}{M^2}\cdot
}
\end{array}
%\right.
%\label{eqCGL35Phiprime}
\label{eqCGL35Order3}
\end{eqnarray}
After a solution $M$ of (\ref{eqCGL35Order3})${}_1$ has been obtained,
the value of $\varphi'$ follows  from (\ref{eqCGL35Order3})${}_3$ 
and the complex amplitude $\GLA$ from (\ref{eqCGL35ReducMphi}).
\medskip

Therefore, we do not need here the structure of singularities of the CGL PDE,
established in \cite{CT1989} (CGL3) and \cite{MCC1994} (CGL5),
we only need the structure of singularities of the third order ODE (\ref{eqCGL35Order3}).
Let us first recall that its solution $G=0$ must be discarded
since it is not a solution of the original system.
Indeed,
the direct substitution of 
$A=\sqrt{M(x-c t)} e^{-i \omega t + i K (x-c t)}$ in CGL3/5
immediately yields $e_i=0$ (CGL5 case) or $d_i=0$ (CGL3 case),
which we explicitly discard as said above.
Under the assumption made in the Introduction ($q/p$ not real if CGL3, $r/p$ not real if CGL5),
this third order ODE evidently fails the Painlev\'e test \cite{CMBook2} since CGL is chaotic,
and,
near a movable singularity $\xi_0$
(which we set to zero because of the invariance under translation),
it admits exactly two Laurent series for CGL3 \cite{MC2003},
\begin{eqnarray}
& & {\hskip -5.0 truemm}
%\hbox{(CGL3) }
M=A_0^2 \xi^{-2}
\left[
1+  \frac{\csi}{3} \xi 
+  \frac{(5 \alpha^2-1)\csi^2 + 12 g_i +24 \alpha g_r}{36 (1+3 \alpha^2)} \xi^2 
+\mathcal{O}(\xi^3)
\right],
\label{eqCGL3-M-series}
\end{eqnarray}
and four Laurent series for CGL5
\cite[Eq.~(21)]{ConteNgCGL5_ACAP}
\cite[Eq.~(18)]{ConteNgCGL5_TMP} 
\begin{eqnarray}
& & {\hskip -15.0 truemm}
%\hbox{(CGL5) }
M=A_0^2 \xi^{-1}
\left[
1+\left(\frac{\csi}{4}+\frac{2 d_r A_0^2-2 e_i d_i A_0^6} {4(1+4\alpha^2)}\right) \xi
+\mathcal{O}(\xi^2)
\right],
\label{eqCGL5-M-series}
\end{eqnarray}
in which the pair $(A_0^2,\alpha)$ of real constants takes two (CGL3) or four (CGL5) values
\cite{CT1989,MCC1994}, 
\begin{eqnarray}
& & {\hskip -15.0 truemm}
\hbox{(CGL3) }
%\left\lbrace
%\begin{array}{ll}
%\displaystyle{
(-1+i \alpha) (-2+i \alpha) p + A_0^2 q=0,
%}\\ \displaystyle{
\alpha^2-3 \frac{d_r}{d_i} \alpha -2=0, A_0^2=\frac{3 \alpha}{d_i}, d_i \not=0,
%}
%\end{array}
%\right.
\label{eqCGL3LeadingOrderComplex}
\end{eqnarray}
\begin{eqnarray}
& & {\hskip -13.0 truemm}
\hbox{(CGL5) }
%\left\lbrace
%\begin{array}{ll}
%\displaystyle{
\left(-\frac{1}{2}+i \alpha\right) \left(-\frac{3}{2}+i \alpha\right) p + A_0^4 r=0,\
%}\\ \displaystyle{
\alpha^2 -2 \frac{e_r}{e_i} \alpha -\frac{3}{4}=0,
A_0^4=\frac{2 \alpha}{e_i}, e_i \not=0.
%}
%\end{array}
%\right.
\label{eqCGL5LeadingOrderComplex}
\end{eqnarray}

% ------------------------------------------------------------------------
\section{An exhaustive method}
\label{sectionNecessary}

The necessary method previously presented in Refs.~\cite{MC2003,CM2009}
and recalled in the present section stems from a quite simple observation:
for all solutions found by the methods of section \ref{sectionSufficient},
the only movable singularities of $M(\xi)$ are poles,
in the complex plane $\mathbb{C}$ of course.
Conversely, let us make the \textit{single} assumption
that all the movable singularities of $M(\xi)$ are poles
(i.e.~that $M$ is meromorphic on $\mathbb{C}$).

Given this assumption, the method which allows one to find all traveling waves
meromorphic on $\mathbb{C}$ relies on the following past achievements.

\begin{enumerate}
\item 
The characterization, by Briot and Bouquet \cite{BriotBouquet},
of all first order autonomous ODEs having a singlevalued general solution
by a privileged class of functions,
made of elliptic functions and their successive degeneracies
(rational functions of one exponential $e^{k \xi}$, rational functions).
	
\item 
The generalization, by Hermite \cite{Hermite-sum-zeta}, 
to elliptic functions and their degeneracies
of the well known partial fraction decomposition of a rational function 
as the sum of a polar part and an entire part.
See details in \cite[Appendix C]{CMBook2}.
	
\item 
A theorem of Eremenko \cite{EremenkoKS} based on Nevanlinna theory \cite{LaineBook}
proving that all meromorphic solutions of a wide class of autonomous algebraic ODEs
are necessarily elliptic or degenerate elliptic
and that the number of such solutions is finite.

\item 
The construction, by two of us \cite{MC2003,CM2009},
of a method (subequation method) to find a closed form expression 
of all elliptic and degenerate elliptic solutions of any algebraic ODE,
however without knowing an upper bound on the number of cases to examine.
		
\item
The proof, by two of us \cite{ConteNgCGL5_ACAP},
that the third order ODE 
for the squared modulus $M$ of CGL3/5 belongs to the class of Eremenko.
\end{enumerate}

In order to make the paper self-contained, let us recall here the last three results.

% ------------------------------------------------------------------------
\subsection{Theorem of Eremenko}
\label{sectionEremenko}

Consider an algebraic autonomous
ODE, i.e.~$P(u^{(N)}(x),\dots,u'(x),u(x))=0$,
in which $P$ is a polynomial of all its arguments.
Eremenko split this class of ODEs in two subclasses,
by proving the following theorem.

\begin{theorem}
(Eremenko \cite{EremenkoKS}).
If an algebraic autonomous ODE enjoys the two properties
(i) it has only one term of maximal global degree in all the derivatives of $u(x)$
(in short, one top degree term);
(ii) the number of its distinct Laurent series (excluding Taylor) is finite,
then any solution meromorphic on $\mathbb{C}$ 
is necessarily elliptic or degenerate elliptic
(i.e.~rational in one exponential $e^{k x}$ or rational in $x$).
\label{Theorem-Eremenko}
\end{theorem}
\smallskip
For a detailed proof, see either \cite{ConteNgWong} 
or \cite[\S 3.2.4]{CMBook2}.
\newline
Let us give one example in each subclass.
\newline
An example which matches the two properties
is the one chosen by Eremenko \cite{EremenkoKS},
the Kuramoto-Sivashinsky ODE,
\begin{eqnarray}
& &
 \nu u''' + b u'' + \mu u' + \frac{u^2}{2} + K = 0,
\label{eqKSODE}
\end{eqnarray}
with $\nu,b,\mu,K$ constant.
The global degree of the five terms is $1,1,1,2,0$,
therefore $u^2/2$ is the unique top degree term.
This ODE only admits movable triple poles, and
the Laurent series near a pole $x=x_0$
(we set $x_0=0$ because of the autonomous nature of the ODE), 
\begin{eqnarray}
{\hskip -12.0 truemm}
& &
u = 120 \nu x^{-3} - 15b x^{-2}
        + \frac{15 (16 \mu \nu - b^2)}{4 \times 19 \nu} x^{-1}
        + \frac{13 (4  \mu \nu - b^2) b}{32 \times 19 \nu^2}
        + O(x),
\label{eqKSODELaurent}
\end{eqnarray}
contains no arbitrary coefficient,
therefore the number of Laurent series is just one.
Once computed \cite{EremenkoKS}, the only meromorphic solutions
are those of Table \ref{TableKS}.

\begin{table}[ht] % [p]
\caption[Kuramoto-Sivashinsky equation. All solutions meromorphic on $\mathbb{C}$.]{
All solutions of KS, Eq.~(\ref{eqKSODE}),   which are meromorphic on $\mathbb{C}$.
% Notation $k^2=-2 S$.
The second and third lines are degeneracies of the elliptic solution.
In the last line, $b=\mu=K=0$.
}
\vspace{0.2truecm}
\begin{center}
\begin{tabular}{| l | c | c | c |}  
\hline 
type    &codimension&$b^2/(\mu\nu)$&$\nu K/\mu^3$\\ \hline \hline 
elliptic&1          & $16$         & arbitrary   \\ \hline 
trigo   &2          & $16$         &$-18$        \\ \hline 
trigo   &2          & $16$         &$-8$         \\ \hline 
trigo   &2          & $144/47$     &$-1800/47^3$ \\ \hline 
trigo   &2          & $256/73$     &$-4050/73^3$ \\ \hline 
trigo   &2          & $0$          &$-4950/19^3$ \\ \hline 
trigo   &2          & $0$          &$450/19^3$   \\ \hline 
rational&3          & $0$          &$0$          \\ \hline
\end{tabular}
\end{center}
\label{TableKS}
\end{table}

An example in the second subclass is
\begin{eqnarray}
& & u'' + 3 u u' +u^3 + a u +b=0,
\end{eqnarray}
which matches the first property (a single top degree term, $u^3$)
but not the second since one of its Laurent series,
\begin{eqnarray}
& & u=x^{-1} + c_1 + (-c_1^2-a/3) x + O(x^2)
\end{eqnarray}
contains the arbitrary coefficient $c_1$,
making the number of Laurent series infinite.
Indeed, the linearizing transformation 
$u=\varphi'/\varphi, \varphi^{'''} + a \varphi'+b \varphi=0$
expresses the general solution as a rational function of two different exponential functions.

% ------------------------------------------------------------------------
\subsection{Subequation method}
\label{sectionSubequation}

It relies on a classical theorem of Briot and Bouquet.
\begin{theorem} \cite[theorem XVII p.~277]{BriotBouquet}.
Given two elliptic functions $u,v$ with the same periods
of respective elliptic orders $m,n$
(i.e.~numbers of poles in a period parallelogram, counting multiplicity),
they are linked by an algebraic equation
\begin{eqnarray}
& &
F(u,v) \equiv
 \sum_{k=0}^{m} \sum_{j=0}^{n} a_{j,k} u^j v^k=0,\ 
a_{j,k} \hbox{ constant},
\label{eqTwoEllFunctions}
\end{eqnarray}
with $\Degree(F,u)=\Order(v)$, $\Degree(F,v)=\Order(u)$.
If in particular $v$ is the derivative of $u$,
the first order ODE obeyed by $u$ takes the precise form
\begin{eqnarray}
& &
F(u,u') \equiv
 \sum_{k=0}^{m} \sum_{j=0}^{2m-2k} a_{j,k} u^j {u'}^k=0,\ a_{0,m}=1.
\label{eqsubeqODEOrderOnePP}
\end{eqnarray}
\end{theorem}
Then, given some algebraic autonomous ODE of any order $N$,
which may admit elliptic solutions,
such as (\ref{eqKSODE}),
the successive steps to obtain all such solutions are
\cite{MC2003,CM2009} (we skip here some unessential details):
\begin{enumerate}

\item
Enumerate all Laurent (not Taylor) series of the $N$-th order ODE,
\begin{eqnarray}
& &
u=x^p \sum_{j=0}^{+\infty} u_j x^j,\ -p \in \mathbb{N}.
\label{eqLaurent}
\end{eqnarray}

\item
For all subsets (including the empty set) of the set of Laurent series,
perform the remaining steps.

\item
Compute the sums $m=\sum p$ and $n=\sum (p+1)$
of the pole orders of the Laurent series of $u$ and $u'$ in the current subset,
and define the first order equation $F(u,u')=0$ (the subequation),
\begin{eqnarray}
& &
F(u,u') \equiv
  \sum_{k=0}^{m} \sum_{j=0}^{n}     a_{j,k} u^j {u'}^k=0,\ 
a_{0,m}=1.
\label{eqsubeqODEOrderOne-mn}
\end{eqnarray}

\item
Compute enough terms $\jmax$ in each Laurent series,
with $\jmax$ slightly greater than the maximal number $(m+1)^2$ 
of coefficients $a_{j,k}$ in (\ref{eqsubeqODEOrderOne-mn}).

\item
Require each Laurent series (\ref{eqLaurent}) 
%[NDLR and the INFINITE Laurent series representing the entire part]
of the current subset 
to obey the subequation $F(u,u')=0$,
\begin{eqnarray}
& & {\hskip -10.0 truemm}
F \equiv x^{m(p-1)} \left(\sum_{j=0}^{\jmax} F_j x^j
 + {\mathcal O}(x^{\jmax+1})
\right),\
\forall j\ : \ F_j=0.
\label{eqLinearSystemFj}
\end{eqnarray}
and solve this \textbf{linear overdetermined} system for $a_{j,k}$.
\end{enumerate}

In the above example (\ref{eqKSODE}) (one triple pole for $u$,
one quadruple pole for $u'$),
only one subequation needs be considered,
\begin{eqnarray}
& & {\hskip -9.0truemm}
%F\equiv 
{u'}^3 
+ (a_{02} +a_{12} u) {u'}^2 
+ (a_{01} +a_{11} u +a_{21} u^2) u' 
+ (a_{00} +a_{10} u +a_{20} u^2 +a_{30} u^3 +a_{40} u^4) 
=0,
\end{eqnarray}
and it is sufficient to stop the series at $\jmax=16$ to find all the 8 subequations.

Each subequation (a first order ODE) is then integrated 
by any method, such as:
the Hermite decomposition,
the computer algebra package \verb+algcurves+ of Maple \cite{vanHoeij-algorithm,MapleAlgcurves},
or other \cite[Chap.~IV]{BriotBouquet}). 

%------------------------------------------------------------------------
\subsection{A property of CGL3/5}
\label{sectionThmCGL35}

As seen in section \ref{sectionSingularities},
the third order ODE (\ref{eqCGL35Order3})${}_1$ for $M(\xi)$ 
possesses the second property of Eremenko's theorem
(finiteness of the number of Laurent series,
two for CGL3, four for CGL5).
As to the first property (one top degree term),
it is true at least for $e_r\not=0$,
the single term being $-4 e_r e_i^2 M^{10}$.
In order to remove this restriction on $e_r$,
the strategy adopted in \cite{ConteNgCGL5_ACAP} is different
and consists in proving that, if $M(\xi)$ is meromorphic and not rational,
firstly it must have infinitely many poles,
secondly it must be periodic.
The precise statement is then

\begin{theorem} \cite[Theorem I p.~156]{ConteNgCGL5_ACAP}. 
For all values of the CGL parameters $p,q,r$ (complex), $\gamma$ (real)
and of the traveling waves parameters $c,\omega$ (real),
for both CGL3 ($q/p$ not real) 
and      CGL5 ($r/p$ not real),
all solutions $M(\xi)$ meromorphic on $\mathbb{C}$ are elliptic or degenerate elliptic
and therefore obey a nonlinear ODE of first order
whose degree is at most two (CGL3) or four (CGL5).
\end{theorem}

%------------------------------------------------------------------------
\subsection{The method}
\label{sectionMethod}

Given the above mentioned preliminary results,
the successive steps to build the complex amplitude $A(x,t)$
of all meromorphic traveling waves of CGL3/5 are then the following.

\begin{enumerate}
\item
Determine all first order subequations for $M(\xi)$
of degree at most two (CGL3) and four (CGL5).

\item
Integrate each subequation by any method (Hermite decomposition,
Maple \verb+algcurves+ package \cite{vanHoeij-algorithm,MapleAlgcurves}
or other \cite[Chap.~IV]{BriotBouquet}). 

\item
For each such expression $M(\xi)$,
compute the logarithmic derivative $(\log \GLa)'$ by the formula 
\begin{eqnarray}
& & {\hskip -13.0 truemm}
\varphi'=\hbox{(\ref{eqCGL35Order3})}_3,
(\log \GLa)'=\frac{M'}{2 M} + i \varphi', 
\label{eqdlogaofM}
\end{eqnarray}
and establish its Hermite decomposition.

\item
Compute the logarithmic primitive $\GLa$ of this Hermite decomposition,
and therefore $A$,
as a product of complex powers of entire functions $\sigma(\xi-\xi_j)$
of Weierstrass \cite[Chap.~18]{AbramowitzStegun} 
or its degeneracies $(2/k) \sinh(k((\xi-\xi_j)/2)$ and $\xi-\xi_j$.
\end{enumerate}
Since every meromorphic solution $M$ can be characterized
either by its Hermite decomposition 
or by its first order equation (the ``subequation''),
it is advisable to combine these two representations
in order to overcome the sometimes heavy computations involved.
In particular, in the worst case 
(four simple poles of $M$ for CGL5 and $\csi=0$),
it proves technically more efficient to compute the subequation first.

Finally, 
using elliptic or trigonometric identities,
the obtained mathematical expression $A$ 
is displayed as a physically relevant formula,
i.e.~$M(\xi)$ bounded for $\xi$ real
and $A$ exhibiting the desired properties (homoclinic or heteroclinic,
front or pulse or source/sink, defect, etc).

% ------------------------------------------------------------------------
\section{Application of the exhaustive method to CGL}
\label{sectionNecessaryCGL}

The subequation is defined as 
the most general autonomous first order ODE with $n$ poles 
of the same order $\poleorder$
whose general solution can be elliptic or degenerate elliptic 
\cite[theorem XVII p.~277]{BriotBouquet},
\begin{eqnarray}
& &
F(M,M') \equiv
\sum_{k=0}^{n \poleorder} \sum_{j=0}^{n(\poleorder+1)-2k} a_{j,k} M^j {M'}^k=0,
 j \poleorder + k(\poleorder+1) \le n \poleorder (\poleorder+1), a_{0,n}=1.
%\sum_{k=0}^{n} \sum_{j=0}^{2n-2k} a_{j,k} M^j {M'}^k=0, a_{0,n}\not=0.
%a_{0,n \poleorder} {M'}^{n \poleorder} + \dots +  a_{n(\poleorder+1),0} M^{n(\poleorder+1)}=0, a_{n(\poleorder+1),0} \not=0.
%\label{eqsubeqODEOrderOnePP}
\end{eqnarray}
The selection rule on $(j,k)$ states that no term can have a singularity degree
higher than that of ${M'}^{n \poleorder}$.

Only six such subequations need to be established:
for CGL3, $\poleorder=2$ and $n=1,2$;
for CGL5, $\poleorder=1$ and $n=1,2,3,4$.
Moreover, the case $n=2$ of CGL5 splits into two subcases 
when one requires the two residues $A_0^2$ to solve the terms of highest singularity degree 
\begin{eqnarray} 
& & M=A_0^2 \xi^{-1}:\ 
a_{40} \left(A_0^2 \xi^{-1}\right)^4 + a_{21} \left(A_0^2 \xi^{-1}\right)^2 \left(-A_0^2 \xi^{-2}\right) + a_{02} \left(-A_0^2 \xi^{-2}\right)^2=0,
\end{eqnarray}
because the two values of $A_0^2$ can be either opposite or nonopposite, see (\ref{eqCGL5LeadingOrderComplex}).

Practically, the computation splits into two successive phases.
The first one is the resolution of a \textit{linear} algebraic system in the unknowns $a_{j,k}$,
this is quick and easy, see e.g.~\cite{MC2003}.
The second phase is the resolution of a \textit{nonlinear} algebraic system 
in the real parameters $d_r,d_i,e_r,e_i,\csi,g_r,g_i$ ($\csr$ drops out)
and the complex locations of the poles;
since the Groebner package of Maple fails to solve most of these nonlinear systems,
one has to do it ``by hand'',
i.e.~to choose which variables to eliminate in order to factorize some equations
into smaller equations
(see e.g.~\cite[\S3.3.9.3]{CMBook2}),
a process which is time, storage and effort consuming.

Finally, the method generates eleven solutions: four of CGL3 and seven of CGL5,
see Table \ref{TableMeroCGL}.

Among them, 
five had never been found by previous methods, they are presented in next sections.
These are precisely all those solutions whose number of poles is maximal.
For completeness, the six others are recalled in an Appendix.
For each solution, 
the displayed information is:
the first order autonomous ODE for $M(\xi)$,
its solution,
two expressions for the complex amplitude.
The first one, which arises from the algorithm,
is a product of complex powers of entire functions $\sigma(\xi-\xi_j)$
or its degeneracies $(k/2)\sinh(k((\xi-\xi_j)/2)$ and $\xi-\xi_j$.
The second one
is the product of a positive modulus by a phase factor of modulus unity,
written so as to display the physical nature of the solution
(homoclinic or heteroclinic, pulse, front, sink, etc).

%\vfill\eject
% ------------------------------------------------------------------------
\subsection{CGL3 elliptic solution}
\label{sectionCGL3Elliptic}

Nongenerate elliptic solutions are easier to determine, for two reasons.
As shown by Briot and Bouquet \cite[\S 181 p.~278]{BriotBouquet},
their first order ODE cannot contain the power one of $M'$, 
which excludes the value $k=1$ in (\ref{eqsubeqODEOrderOnePP}).
The second reason is the necessary condition that 
the sum of the residues of the Laurent series (\ref{eqCGL3-M-series})
of $M$ 
(or more generally of any power of a derivative of $M$ of any order)
inside a period parallelogram vanishes.
Assuming $M$ to have only one pole yields no elliptic solution,
but, with two poles,
$M$   generates the condition $d_r \csi=0$,
$M^2$           the condition $(\csi^2 + 6 g_i) \csi=0$.
Indeed, the two constraints $d_r=0, g_i=-(1/6) \csi^2$
do generate a unique elliptic solution.
Its particular case $\csi=0$ is also elliptic.

This solution occurs
for the exponent $\alpha=\pm \sqrt{2}$,
% \verb+sub32eM+ 
\begin{eqnarray}
& & {\hskip -15.0 truemm}
\hbox{(CGL3) }
\left\lbrace
\begin{array}{ll}
\displaystyle{
d_r=0,\ g_i=-\frac{1}{6} \csi^2,\ 
}\\ \displaystyle{
 3^7\ 7^6 (d_i M')^4 
-2^3\ 3^5\ 7^5\  \csi (7 d_i M - 2 g_r)(d_i M')^3
}\\ \displaystyle{
+2\ 3^2\ 7^4 \csi^2
\left[18 (7 d_i M - g_r) (35 d_i M -17 g_r) -49 \csi^4\right](d_i M')^2
}\\ \displaystyle{
-2^3\ 3^4\ ((7 d_i M)^2 - 56 g_r d_i M -2 g_r^2)^3
}\\ \displaystyle{
+ 7^2 \csi^4
\left[
(-3 (7 d_i M)^2 + 2^6\ 7 g_r d_i M +66 g_r^2 +49 \csi^4)^2
\right. 
}\\ \displaystyle{
\left. \phantom{123456}
-7 ( 147 (d_i M)^2 + 28 g_r d_i M +36 g_r^2)
   ( 441 (d_i M)^2 -308 g_r d_i M +24 g_r^2)
\right]=0.
}
\end{array}
\right.
\label{eqCGL3Ellipsubeq}
\end{eqnarray}

\textit{Remark}.
The restrictive assumption $d_r=1/2$ made in Refs.~\cite{MC2003} \cite{Hone2005}
prevented this elliptic solution to be found earlier.
The reason why it was also not detected in Ref.~\cite{VernovCGL3}
is different:
since the number of Laurent series of $\psi$ is infinite 
because of the presence of an arbitrary coefficient
(see e.g.~\cite[Eq.~(21)]{ConteNgCGL5_TMP}),
the function $\psi$ should not be used 
to build a subequation for $\psi(\xi)$.

After scaling, the solution of (\ref{eqCGL3Ellipsubeq})
depends on a single parameter $g_r/\csi^2$,
unless $\csi$ vanishes,
in which case the subequation has the binomial type,
\begin{eqnarray}
& & {\hskip -15.0 truemm}
d_r=g_i=\csi=0 :\ 
(d_i M')^4 - \frac{8}{9} \left((d_i M)^2 - 8 \frac{g_r}{7} d_i M -2 \left(\frac{g_r}{7}\right)^2\right)^3=0.
\end{eqnarray}

The simplest way to integrate (\ref{eqCGL3Ellipsubeq}) is to represent $M$
by its Hermite decomposition in which one of the two poles is put at the origin,
\begin{eqnarray}
& & {\hskip -15.0 truemm}
M=\frac{3 \sqrt{2}}{d_i} 
\left\lbrack\wp(\xi) -\wp(\xi-\xi_a) + \frac{\csi}{3} \left(\zeta(\xi)-\zeta(\xi-\xi_a) - \zeta(\xi_a) \right)
+ c_0 \right\rbrack.
\label{eqCGL3EllipM0Hermite}
\end{eqnarray}
Indeed, the coefficients of the two polar parts are known,
these are those of the two Laurent series (\ref{eqCGL3-M-series}) for $\alpha=\pm \sqrt{2}$.
The technique \cite{DK2011method} to determine
the constants $c_0,\wp(\xi_a),\wp'(\xi_a),g_2,g_3$ 
is to identify the different Laurent series with the expansions of $M$ 
near the various poles (here $0,\xi_a$).
The result is % \verb+M0dis=M0disrat+,
% invariance \sqrt{2} \to -\sqrt{2} by permutation of the two poles
\begin{eqnarray}
& & {\hskip -15.0 truemm}
%\hbox{(CGL3) }
\left\lbrace
\begin{array}{ll}
\displaystyle{
%rulea:={WeierstrassP(a,wg2,wg3)=-(csi/6)**2,WeierstrassPPrime(a,wg2,wg3)=sqrt(2)*ogr*csi/126};
%Rulea:={wp0a                   =-(csi/6)**2,wp1a                        =sqrt(2)*ogr*csi/126};
%M0dis:=(3*sqrt(2)/di)*(WeierstrassP(xi,wg2,wg3)
%-WeierstrassP(xi-a,wg2,wg3)
%  +(csi/3)*(WeierstrassZeta(xi,wg2,wg3)-WeierstrassZeta(xi-a,wg2,wg3)-WeierstrassZeta(a,wg2,wg3))
%	+(2*sqrt(2)*ogr)/21);
%
%\hbox{M0dis}=
M=\frac{3 \sqrt{2}}{d_i} 
\left\lbrack\wp(\xi) -\wp(\xi-\xi_a) + \frac{\csi}{3} \left(\zeta(\xi)-  \zeta(\xi-\xi_a) - \zeta(\xi_a) \right)
+\frac{2 \sqrt{2} g_r}{21} \right\rbrack
}\\ \displaystyle{
\phantom{M}
% 3*sqrt(2)*wp0+(4/7)*ogr+(1/12)*sqrt(2)*csi^2
% +(1/49)*(27*sqrt(2)*ogr^2-882*sqrt(2)*csi*wp1-14*csi^2*ogr)/(csi^2+36*wp0)
% -(12/49)*csi*ogr*(sqrt(2)*ogr*csi+126*wp1)/(csi^2+36*wp0)^2
=\frac{1}{d_i} \left[
3 \sqrt{2} \wp(\xi) + \frac{4}{7} g_r 
+ \frac{\sqrt{2}}{12} \csi^2
+ \frac{27 \sqrt{2} g_r - 14 \csi^2 g_r - 882 \sqrt{2} \csi \wp'(\xi)}{49 (36 \wp(\xi)+\csi^2)} 
\right.
}\\ \displaystyle{
\phantom{123456789012345678901234567890}
\left.
- \frac{12 \csi g_r (\sqrt{2} g_r \csi + 126 \wp'(\xi))}{49 (36 \wp(\xi)+\csi^2)^2} 
\right],
}\\ \displaystyle{
\wp (\xi_a)=-\left(\frac{\csi}{6}\right)^2,\ 
\wp'(\xi_a)= \frac{\sqrt{2}}{126}\csi g_r,\
}\\ \displaystyle{
g_2=\frac{g_r^2}{49}+\frac{\csi^4}{108}\ccomma
g_3=\left(\frac{g_r^2}{7}+\frac{\csi^4}{18}\right) \frac{\csi^2}{324}\ccomma
}\\ \displaystyle{
%}\\ \displaystyle{
g_2^3-27 g_3^2=2^{-4} 3^{-7} 7^{-6} g_r^2 (243 g_r^2+98 \csi^4)(144 g_r^2+49\csi^4).
%}\\ \displaystyle{
%4 z^3-g_2 z -g_3=\hbox{not factorizable on the extension } \mathbb{Q}, i, \sqrt{2}.
}
\end{array}
\right.
\label{eqCGL3EllipM0dis}
\end{eqnarray}

The only degeneracy to a simply periodic solution occurs for $g_r=0$, 
in which case the subequation is decomposable,
\begin{eqnarray}
& & {\hskip -15.0 truemm}
g_r=0:\
%(3 d_i^2 (9 {M'}^2 - 12 \csi  M M' + 2 \csi^2 M^2) + \csi^6)^2
% -18 d_i^2 (2 \csi^3 M'-6 d_i^2 M^3 - \csi^4)^2=0,
3 d_i^2 (9 {M'}^2 - 12 \csi  M M' + 2 \csi^2 M^2) + \csi^6
 \pm 3 \sqrt{2} d_i (2 \csi^3 M'-6 d_i^2 M^3 - \csi^4)=0,
\label{eqCGL3Ellipgr0}
\end{eqnarray}
and represents the propagating hole (\ref{eqCGL3Hole}) % \cite{BN1985}
for the particular values $\alpha=\pm \sqrt{2}, g_r=0$.

When $g_r$ is nonzero, 
the remaining question is to determine whether the elliptic expression (\ref{eqCGL3EllipM0dis})
represents a real bounded solution or not.
This single complex expression depends on the (omitted) arbitrary complex origin $\xi_0$ of $\xi$,
therefore it represents in fact two real solutions,
one for $\xi_0=0$ and one for $\xi_0$ equal to the nonreal half-period
(exactly like $\tanh(\xi-\xi_0)$  represents both the bounded front $\tanh(\xi)$
and the unbounded real function $\coth(\xi)$).
In the present case, none of the two real solutions is bounded on the real axis $\xi$
(see Figures 3.4 and 3.5 in \cite{CMBook2}),
therefore the solution is physically meaningful only
for its degeneracy $g_r=0$ to the traveling hole.

The expression of the complex amplitude $A$ as a product of powers of $\sigma$ functions
can be found in \cite[Eq.~(3.181)]{CMBook2}.

%\vfill\eject
% ------------------------------------------------------------------------
\subsection{CGL5 elliptic solution}
\label{sectionCGL5Elliptic}

In order to detect a CGL5 elliptic solution,
the criterium of residues 
needs to be applied not only to $M$ but also \cite{ConteNgCGL5_TMP} to
$\left(M^{(k)}\right)^j$,
$(j,k)=(0,2),(0,3),(0,4),(1,2)$,
it shows that $M$ must possess four poles
and the parameters must obey four real constraints.

Only one subequation exists, for $\alpha=\pm \sqrt{3}/2$
\cite{ConteNgCGL5_ACAP} 
\cite{ConteNgCGL5_TMP},
\begin{eqnarray}
\left\lbrace
\begin{array}{ll}
\displaystyle{
e_r=d_r=d_i=0,\ g_i=-\frac{3}{16} \csi^2,\ 
}\\ \displaystyle{
% Maple subeq4m2  file cgl5.Laurent4.elliptic.subeqpsi.mw
% AMP   sub54ellM = sub54eM file nls.amp
e_i^2 {M'}^4 
-2 \csi e_i^2 M {M'}^3
 +\frac{1}{2} \csi^2 (3 e_i M^2 -g_r) e_i {M'}^2
- \frac{1}{3^4}  e_i M^2 (3 e_i M^2 - 4 g_r)^3
}\\ \displaystyle{\phantom{12345}
 + \frac{1}{32} \csi^4 (- 9 e_i^2 M^4 + 6 g_r e_i M^2 + 2 g_r^2)
 +\frac{3^4}{2^{12}} \csi^8
=0.
%}\\ \displaystyle{
%{M'}^4 
%-2 \csi M {M'}^3
% +\frac{72}{e_i} \ex {M'}^2 (e_i M^2 - 12 \ey)
% +\frac{2^4 3^8 \ex^4}{e_i^2}
%}\\ \displaystyle{\phantom{1234}
% +\frac{648 \ex^2}{e_i^2} \left(288 \ey^2 +24 e_i \ey M^2 -e_i^2 M^4\right)
% - \frac{1}{3^4 e_i} M^2 \left(e_i M^2 -48 \ey\right)^3
%=0,
%\label{eqsub54Mexey}
}
\end{array}
\right.
\label{eqCGL5Ellipsubeq}
\end{eqnarray}

The particular case $\csi=0$ of this solution was first found by Vernov \cite{VernovCGL5},
who obtained two binomial subequations of the type of Briot and Bouquet,
\begin{eqnarray}
& & e_r=d_r=d_i=g_i=\csi=0, g_r\not=0,\
\left\lbrace
\begin{array}{ll}
\displaystyle{
e_i^2 (3 M')^4 - e_i M^2 \left(3 e_i M^2 -4 g_r\right)^3=0,
}\\ \displaystyle{
9 {\psi'}^2 -12  \psi^4 - g_r^2=0, \psi=\varphi'-\csr/2.
}
\end{array}
\right.
\label{eqsubVernov}
\end{eqnarray}
The reason why he did not find the full subequation 
is his use of $\psi$ instead of $M$:
since there exists a Laurent series of $\psi$ with an arbitrary coefficient,
the number of Laurent series of $\psi$ is infinite and
Eremenko's theorem does not apply to the third order ODE for $\psi$.

Integrating (\ref{eqCGL5Ellipsubeq}) as a rational function of $\wp$ and $\wp'$
\cite[\S 18.11]{AbramowitzStegun},
as done in \cite[Eq.~(46)]{ConteNgCGL5_ACAP},
creates useless complications (Landen transformation, etc),
so let us shortly describe the good procedure.

Let us for convenience define the notation
\begin{eqnarray}
& & 
%e_i=\sqrt{3}/a_2^2, 
\ex=\frac{\csi^2}{48},
\ey=\frac{g_r}{36}\cdot
\end{eqnarray}

Assuming a Hermite decomposition with one of the four poles at the origin,
another pole ($\xi_2$) is real 
and the two others ($\xi_1,\xi_3$) are complex conjugate,
\begin{eqnarray}
& & 
\left\lbrace
\begin{array}{ll}
\displaystyle{
% AMP M0Hermite
\frac{M}{A_0^2}=\frac{\csi}{4}
+ \zeta(\xi,g_2,g_3) + \sum_{k=1}^3
  i^{k} \left(\zeta(\xi-\xi_{k},g_2,g_3) +\zeta(\xi_{k},g_2,g_3)\right), i^2=-1,
%\csi^2=48 \ex,\ g_r=36 \ey,\ e_r=d_r=d_i=0,\ g_i=-\frac{3}{16} \csi^2. 
\label{eqCGL5Ellip-M-Assumption}
}\\ \displaystyle{
% ======================================== M/a_2=rational(wp,wp',\csi,g_r)
%\frac{M}{a_2}=\frac{\csi}{4}
% +\frac{-\frac{\csi}{32} (\wp(\xi)-\frac{\csi^2}{48}) \left[\csi^2 \wp(\xi)+\frac{\csi^4}{24}+\frac{4}{27} g_r^2\right] 
% + \wp' \left[-\frac{1}{2} (\wp(\xi)+\frac{\csi^2}{24} + \frac{\sqrt{3}}{18} g_r)^2 + \frac{\sqrt{3}}{288} g_r \csi^2 \right]}
%  {(\wp(\xi)-\frac{\csi^2}{48})\left[\left(\wp(\xi)+\frac{\csi^2}{24}\right)^2 + \frac{g_r^2}{108}  \right]}\ccomma
%}\\ \displaystyle{
% ======================================== M/a_2=rational(wp,wp',\ex,\ey,\csi)
%\frac{M}{a_2}=\frac{\csi}{4}
% +\frac{- \csi (\wp(\xi)-\ex) \left[\frac{3}{2} \ex \wp(\xi)+ 6 \ey^2 + 3 \ex^2 \right] 
% + \wp' \left[-\frac{1}{2} (\wp(\xi)+2 \ex + 2 \sqrt{3} \ey)^2 + 6 \sqrt{3} \ex \ey \right]}
% {(\wp(\xi)-\ex)\left[\left(\wp(\xi)+2 \ex\right)^2 + 12 \ey^2  \right]}\ccomma
%}\\ \displaystyle{
% AMP M0wp
\phantom{\frac{M}{A_0^2}}
 =\frac{\csi}{4}
 - \csi \frac{\frac{3}{2} \ex \wp(\xi)+ 6 \ey^2 + 3 \ex^2} {\left(\wp(\xi)+2 \ex\right)^2 + 12 \ey^2}
 + \wp' \frac{\left[-\frac{1}{2} (\wp(\xi)+2 \ex + 2 \sqrt{3} \ey)^2 + 6 \sqrt{3} \ex \ey \right]}
 {(\wp(\xi)-\ex)\left[\left(\wp(\xi)+2 \ex\right)^2 + 12 \ey^2  \right]}\ccomma
}\\ \displaystyle{
%
%let wp(xi2)=e1,'(wp,1)(xi2)=0 in answer;
%let wp(xi1)=-2*e1+2*i*r3*e0,'(wp,1)(xi1)=csi*( i*3/2*e1+r3*e0) in answer;
%let wp(xi3)=-2*e1-2*i*r3*e0,'(wp,1)(xi3)=csi*(-i*3/2*e1+r3*e0) in answer;
%  ,g_2,g_3
\wp(\xi_2,g_2,g_3)=   \ex,                     \wp'(\xi_2,g_2,g_3)=0, 
}\\ \displaystyle{
\wp(\xi_1,g_2,g_3)=-2 \ex + 2 i \sqrt{3} \ey,  \wp'(\xi_1,g_2,g_3)=\left(\sqrt{3} \ey + i \frac{3}{2} \ex \right) \csi, 
}\\ \displaystyle{
\wp(\xi_3,g_2,g_3)=-2 \ex - 2 i \sqrt{3} \ey,  \wp'(\xi_3,g_2,g_3)=\left(\sqrt{3} \ey - i \frac{3}{2} \ex \right) \csi, 
}\\ \displaystyle{
g_2=-24 (  \ex^2 +  2 \ey^2),\
g_3=  4 (7 \ex^2 + 12 \ey^2) \ex,\
{\wp'}^2=4 (\wp-\ex) (\wp^2 + \ex \wp + 7 \ex^2 + 12 \ey^2),\ 
}\\ \displaystyle{
g_2^3-27 g_3^2
=-2^4\ 3^3(9 \ex^2+16 \ey^2)(3 \ex^2+4 \ey^2)^2
=-\frac{(81 \csi^4+256 g_r^2)(64 g_r^2+27 \csi^4)^2}{2^{20}\ 3^9}\cdot
%
%g_2=-\frac{\csi^4}{96} - \frac{g_r^2}{27}\ccomma
%g_3= \left(\frac{7}{27648} \csi^4 + \frac{g_r^2}{1296}\right) \csi^2,
%}\\ \displaystyle{
%g_2^3-27 g_3^2
%=-\frac{(81 \csi^4+256 g_r^2)(64 g_r^2+27 \csi^4)^2}{2^{20}\ 3^9}\ccomma\
}
\end{array}
\right.
\label{eqCGL5EllipM}
\end{eqnarray}

Again, none of the two associated real solutions for the modulus $M$ is bounded,
therefore the solution is unphysical.
Moreover, the only nonelliptic degeneracy is the rational solution
defined by 
\begin{eqnarray}
& & e_r=d_r=d_i=g_i=\csi=g_r=0, 
%e_i^2 (3 M')^4 - e_i M^2 \left(3 e_i M^2\right)^3=0,
3 {M'}^4 - e_i^2 M^8=0.
\end{eqnarray}

As explained in Section \ref{sectionCGL3Elliptic},
the logarithmic derivative of $\GLA e^{i \omega t}$
is the sum of functions with six simple poles only,
made of the four poles $0,\xi_1,\xi_2,\xi_3$ of $M$
and of two poles $\xi_4,\xi_5$ out of the four zeroes of $M$,
%(from now on, $\wp(*)$ and $\zeta(*)$ are short for $\wp(*,g_2,g_3)$ and $\zeta(*,g_2,g_3)$),
\begin{eqnarray}
& & 
%\left\lbrace
%\begin{array}{ll}
%\displaystyle{
(\log \GLa)'=c_0 
 + \frac{-1+i \sqrt{3}}{2} (\zeta(\xi      )+\zeta(\xi-\xi_2))
 + \frac{-1-i \sqrt{3}}{2} (\zeta(\xi-\xi_1)+\zeta(\xi-\xi_3))
\nonumber\\ & & \phantom{1234567}
 +                         (\zeta(\xi-\xi_4)+\zeta(\xi-\xi_5)).
%}
%\end{array}
%\right.
\label{eqCGL5Ellip-dloga}
\end{eqnarray}

\textit{Remark}.
The expression
% (\ref{eqCGL3EllipdlaHermite}) %RC suppressed
(\ref{eqCGL5Ellip-dloga})
belongs to the class of assumptions
\begin{eqnarray}
& & M=\hbox{regular}+\sum_{j=1}^N \mathcal{D}_j \log \psi(\xi-\xi_j),
\label{eqTruncation}
\end{eqnarray}
in which $N$ is at most equal to the number of poles inside a period parallelogram,
$\psi$ is some entire function
(in the simplest cases the solution of a linear ODE with constant coefficients),
and $\mathcal{D}_j$ is the ``singular part operator''
% any ref? \cite{WTC}?,
i.e.~the linear differential operator which represents
all the polar part of the Laurent series.
Therefore the decomposition devised by Hermite in 1888
can be identified with a ``$N$-family truncation method'' for autonomous algebraic ODEs.

% ======================================================================
\subsection{CGL5 homoclinic defect}

The most involved situation is CGL5 with at the same time four poles,
a genus zero subequation (i.e.~a degenerate elliptic solution) and $\csi=0$.
Indeed, among the nonlinear algebraic equations in the fixed parameters
$\csi,e_r,d_r,d_i,g_r,g_i$ and the movable locations $\xi_j$ of the poles,
many of them contain the factor $\csi$,
which allows one to rapidly discard the case $\csi\not=0$,
and the remaining nonzero equations for $\csi=0$ are much bigger.
After elimination of all variables but $e_r/e_i$,
one obtains ten values of $e_r/e_i$, among them two complex ones (discarded) and
\begin{eqnarray}
& & 
\begin{array}{ll}
\displaystyle{
(2 e_r-3 e_i)(1089 e_i^6-81327 e_r^2 e_i^4+323512 e_r^4 e_i^2 +456976 e_r^6) e_r=0.
%}\\ \displaystyle{
}
%write 0==(1/15)*(pr*qi-pi*qr)**2/(pr*ri-pi*rr)/(pr**2+pi**2)-answer;;
%=\frac{\Im(q/p)^2}{15 \Im(r/p)}, 
\end{array}
\end{eqnarray}
These three factors yield the three solutions now presented.

For $\alpha=3/2 \pm \sqrt{3}$,
there exists one four-pole subequation \cite{CMNW-CGL35-Letter}
\begin{eqnarray}
& & {\hskip -20.0 truemm}
\left\lbrace
\begin{array}{ll}
\displaystyle{
%let rule r5defect
% csi=0,er=(3/2)*ei,dr=(29/15)*di,ogr=-(12/35)*ogi,
% ei=a2**(-2)*(3+2*r3),di=a2**(-1)*(-5)*(9+5*r3)*Td,ogi=(175/6)*(12+7*r3)*Td**2,
% r3**2=3 inactive;
\csi=0,
 \frac{e_r}{e_i}=\frac{3}{2},
 \frac{d_r}{d_i}=\frac{29}{15},
 \frac{g_r}{g_i}=-\frac{12}{35},
      g_i=\frac{7 d_i^2}{12 e_i}\ccomma
\label{eqCGL5Trigo4DefectSubeqParam}
}\\ \displaystyle{	
\left({M'}^2 + e_i M \left( M + \frac{2 d_i}{3 e_i}\right) \left(M^2 + \frac{6 d_i}{5 e_i} M + \frac{d_i^2}{3 e_i^2}\right) \right)^2
 -\frac{4}{3} e_i^2 M^2 \left(M^2 + \frac{6 d_i}{5 e_i} M + \frac{d_i^2}{3 e_i^2}\right)^3=0,
%}\\ \displaystyle{	
%\left({M'}^2 + e_i M \left( M + \frac{2 d_i}{3 e_i}\right) P_2 \right)^2
% -\frac{4}{3} e_i^2 M^2 P_2^3=0,
%}\\ \displaystyle{
%P_2=M^2 + \frac{6 d_i}{5 e_i} M + \frac{d_i^2}{3 e_i^2}\ccomma
}
\end{array}
\right.
%\label{eqCGL5Trigo4DefectSubeq}
%\label{eqCGL5Trigo4h}
\end{eqnarray}

The genus of this subequation $F(M',M)=0$ is zero,
therefore its solution is either trigonometric or rational, 
and its algorithmic integration is made \textit{via} its Hermite decomposition
(here trigonometric),
\begin{eqnarray}
& & 
\begin{array}{ll}
\displaystyle{
M=
\left(\sum_{j=1}^4  c_{j} 
 \left(\frac{k}{2}\coth\frac{k}{2}(\xi-\xi_j)\right)  \right)
+ \left(\sum_{m=M_1}^{M_2} d_m (e^{k \xi})^m\right),\
%k\not=0,
}
\end{array}
\label{eqHermitetrigo}
\end{eqnarray}
in which
$M_1$ and $M_2$ are signed integers defining the entire part,
$c_{j}$, $d_m$, $k$ complex constants.
The entire part 
(to be computed as explained in \cite[\S 3.3.3 example 3.4]{CMBook2})
reduces here to zero,
and the final result is 
a unique real bounded value of $M$, for $e_i=\Im(r/p)>0$,
having a unique minimum $M=0$ at the origin,
making it an exact representation of a defect,
\begin{eqnarray}
& & 
%\begin{array}{ll}
%\displaystyle{
M=- 20 \frac{d_i}{e_i} 
 \frac{ \sinh^2 \displaystyle\frac{k \xi}{2} }
      {8\sinh^4 \displaystyle\frac{k \xi}{2}
		 +36\sinh^2 \displaystyle\frac{k \xi}{2}+3}\ccomma\
			k^2=\frac{d_i^2}{15 e_i},
%}
%write 0==(1/15)*(pr*qi-pi*qr)**2/(pr*ri-pi*rr)/(pr**2+pi**2)-answer;;
%=\frac{\Im(q/p)^2}{15 \Im(r/p)}, 
%\end{array}
\label{eqCGL5Defect}
\end{eqnarray}
see Figure \cite[Fig.~1]{CMNW-CGL35-Letter}.

Depending on $p$, this defect is either ($p$ real) moving with an arbitrary velocity $c$
or ($p$ not real) stationary.

Computing the complex amplitude follows the same logic as in section \ref{sectionCGL3Elliptic}.
Using Eqs.~(\ref{eqdlogaofM}) and (\ref{eqCGL5Defect}),
one successively computes $(\log \GLa)'$ as a rational function of $\coth k (\xi-\xi_0)/2$,
then its Hermite decomposition as a finite sum of $\coth(k(\xi-\xi_0-\xi_j)/2)$ terms,
finally the logarithmic primitive of this decomposition.
One thus represents
$(\log \GLa)'$ as the sum of five $\coth(k(\xi-\xi_0-\xi_j)/2$ terms,
therefore the complex amplitude $\GLA$ is the product of complex powers of
five $\sinh(k(\xi-\xi_0-\xi_j)/2)$ functions
\cite[Eq.~(11)]{CMNW-CGL35-Letter}.
This defect is one more elementary analytic pattern,
it could help to describe the mechanism of turbulence called
defect-mediated turbulence 
\cite{LegaThese} \cite{Lega2001}, 
and
% 2001 \cite{Lega2001} page 271 \P 3: 2-dim CGL3 displays defect-mediated turbulence.
it has indeed been observed in numerical simulations of CGL5 \cite[Fig.~3b]{PSAWK1993} \cite[Fig.~4]{PSAK1995}
\cite[p 278]{Lega2001}.

%\vfill\eject
% ======================================================================
\subsection{CGL5 homoclinic bound state of two dark solitons}

Also at the price of five constraints among the fixed parameters,
% $k^2$, $\Mshift$ and $\coth_A/\coth_B+\coth_B/\coth_A$,
%are polynomials of $\ersurei=e_r/e_i$, % (denoted $\ersurei$), % of degree at most five,
%either even or odd,
\begin{eqnarray}
& & {\hskip -20.0 truemm}
\left\lbrace
\begin{array}{ll}
\displaystyle{
\csi=0,
%1089-81327 \ersurei^2+323512 \ersurei^4+456976 \ersurei^6=0,
}\\ \displaystyle{
\frac{e_r}{e_i}=\ersurei = \hbox{ one of the four real roots of }
1089-81327 \ersurei^2+323512 \ersurei^4+456976 \ersurei^6=0, 
}\\ \displaystyle{
d_r=d_i
\ersurei \frac{828038745921+7649070764998 \ersurei^2+9025535790856 \ersurei^4}{1386644084775}\ccomma
}\\ \displaystyle{
g_r=\frac{d_i^2}{e_i} 
 \frac{-58513290148629717+1113015753503375224 \ersurei^2+1243896610551884848 \ersurei^4}{178728931719095040}\ccomma
}\\ \displaystyle{
g_i=\frac{d_i^2}{e_i} \ersurei
 \frac{119473478956925997-1651180178874084664 \ersurei^2-1567990451264571568 \ersurei^4}{1608560385471855360}\ccomma
}
\end{array}
\right.
\label{eqCGL5-csi0-boundstate-fixed}
\end{eqnarray}
there exists a subequation, % (\ref{eqCGL5Trigo4wSubeq}),
\begin{eqnarray}
& & %{\hskip -10.0 truemm}
%\left\lbrace
\begin{array}{ll}
\displaystyle{
\left({M'}^2 + c_6 P_{2a}(M) P_{2b}(M)\right)^2-c_7 (M + c_1)^2 P_{2a}(M)^3=0,
}\\ \displaystyle{
P_{2a}(M)=M^2+c_2 M + c_3,
P_{2b}(M)=M^2+c_4 M + c_5,
}
\end{array}
%\right.
%\label{eqCGL5Trigo4wSubeq}
\end{eqnarray}
where the $P_n$'s are polynomials of degree $n$
and the constants $c_j$ are polynomial in $\ersurei, e_i, d_i$.

To each one of these four real roots $e_r/e_i$,
\begin{eqnarray}
& & 
\begin{array}{ll}
\displaystyle{
%ersurei=\pm 0.119204503224, \alpha=(\mp 0.7549863876, \pm 0.9933953940), %=E_i^2=(2.280017782, 3.947337636),
%ersurei=\pm 0.429992076346, \alpha=(\mp 0.5369066700, \pm 1.396890823),  %E_i^2= (1.153075089,7.805215883), 
\ersurei=\pm 0.1192,       
\ersurei=\pm 0.4300,
}      
\end{array}
\end{eqnarray}
correspond two values $\alpha_1, \alpha_2$
of the exponent $\alpha$ defined in (\ref{eqCGL5LeadingOrderComplex}),
whose product is $-3/4$,
\begin{eqnarray}
& & 
\begin{array}{ll}
\displaystyle{
\ersurei=\frac{e_r}{e_i}=\frac{\alpha}{2}-\frac{3}{8 \alpha},
}\\ \displaystyle{ 
%ersurei=\pm 0.119204503224, \alpha=(\mp 0.7549863876, \pm 0.9933953940), %=E_i^2=(2.280017782, 3.947337636),
\ersurei=\pm 0.1192        , (\alpha_1,\alpha_2)=(\mp 0.7550      , \pm 0.9934      ), %=E_i^2=(2.280017782, 3.947337636),
}\\ \displaystyle{ 
%ersurei=\pm 0.429992076346, \alpha=(\mp 0.5369066700, \pm 1.396890823),  %E_i^2= (1.153075089,7.805215883), 
\ersurei=\pm 0.4300        , (\alpha_1,\alpha_2)=(\mp 0.5369      , \pm 1.397      ),  %E_i^2= (1.153075089,7.805215883), 
}
\end{array}
\end{eqnarray}
and one real bounded modulus, 
\begin{eqnarray}
& & {\hskip -5.0 truemm}
\begin{array}{ll}
\displaystyle{
M
= \Mshift + \frac{d_i}{e_i}
 \frac{\Cthree -(\Cthree+\Cfour) \cosh^2 \displaystyle\frac{k \xi}{2}}
      {1-(2+D_1)                 \cosh^2 \displaystyle\frac{k \xi}{2}
			  +(1+D_1+D_0)             \cosh^4 \displaystyle\frac{k \xi}{2}}\cdot
}
\label{eqCGL5BoundM0sinh}
\end{array}
\end{eqnarray}
Its coefficients are algebraic expressions of $\ersurei=e_r/e_i$,
%\hfill\break\noindent (Maple cgl5.Laurent4.csi0.trigo.Symmetric.mw)
\begin{eqnarray}
& & {\hskip -20.0 truemm}
\left\lbrace
\begin{array}{ll}
\displaystyle{
k^2=\frac{d_i^2}{e_i} \ersurei
 \frac{470354925826628997+15744055491100758536\ersurei^2+16800138410952093392\ersurei^4}{2010700481839819200}\ccomma
}\\ \displaystyle{
}\\ \displaystyle{
\Mshift=\frac{d_i}{e_i}
\frac{-344373082347+2958053216864 \ersurei^2+3382994698928 \ersurei^4}{493029007920}\ccomma
}\\ \displaystyle{	
}\\ \displaystyle{	
% coth22pcoth32normed:=2*(51161905779+18447053072*ersurei^2-65496542176*ersurei^4)/70239007575:									
\coth^2\frac{k (\xi_B-\xi_A)}{2}+\coth^2\frac{k (\xi_B+\xi_A)}{2}=	
2\frac{51161905779+18447053072\ersurei^2-65496542176\ersurei^4}{70239007575}\ccomma
}\\ \displaystyle{		
}\\ \displaystyle{		
% coth22mcoth32normed:=8*I*sqrt(3)*ersurei
%   *(18414161641353-66747311650346*ersurei^2-97033527316232*ersurei^4)/25496759749725:						
\coth^2\frac{k (\xi_B-\xi_A)}{2}-\coth^2\frac{k (\xi_B+\xi_A)}{2}=
8 i \sqrt{3} \ersurei \frac{18414161641353-66747311650346\ersurei^2-97033527316232\ersurei^4}{25496759749725}\ccomma
}\\ \displaystyle{
}\\ \displaystyle{
%\frac{\tau_A^2+\tau_B^2}{\tau_A\tau_B} 
\frac{\coth^2\displaystyle\frac{k \xi_A}{2}+\coth^2\displaystyle\frac{k \xi_B}{2}}
     {\coth  \displaystyle\frac{k \xi_A}{2} \coth  \displaystyle\frac{k \xi_B}{2}}
%\frac{\coth_A}{\coth_B}+\frac{\coth_B}{\coth_A}
%sumratiotauXnormed:=2*I*sqrt(3)*ersurei*(16223643-41722436*ersurei^2-37472032*ersurei^4)/6800175:
 = 2 i \sqrt{3} \ersurei \frac{16223643-41722436 \ersurei^2-37472032 \ersurei^4}{6800175}\ccomma 	
}\\ \displaystyle{
D_1=-\coth^2\frac{k \xi_A}{2}-\coth^2\frac{k \xi_B}{2}\ccomma\ % =\hbox{CjDjnumold(D1)},
%}\\ \displaystyle{
D_0=\coth^2\frac{k \xi_A}{2} \coth^2\frac{k \xi_B}{2}\ccomma % =\hbox{CjDjnumold(D0)},
}\\ \displaystyle{											
-\frac{\Cfour}{\Cthree}=
%cothN2:=(2*alpha*cothB+I*sqrt(3)*cothA)*cothA*cothB/(2*alpha*cothA+I*sqrt(3)*cothB);
\frac{2 \alpha \coth_B+i\sqrt{3} \coth_A}{2 \alpha \coth_A+i\sqrt{3} \coth_B}\coth_A \coth_B\ccomma\
\coth_A=\coth\frac{k \xi_A}{2}, \coth_B=\coth\frac{k \xi_B}{2}, 
}\\ \displaystyle{											
}\\ \displaystyle{											
\Cthree=
% -(1/2)*ei*k*a2*(2*alpha*cothA+I*sqrt(3)*cothB)/(di*alpha);
- \frac{e_i k (2 \alpha \coth_A+i\sqrt{3} \coth_B)}{2 d_i \alpha} \sqrt{\frac{2\alpha}{e_i}}\ccomma
%}\\ \displaystyle{											
%\coth^2\frac{k \xi_N}{2}=\hbox{deducible from the above},
}
\end{array}
\right.
\label{eqCGL5-csi0-boundstate-movable}
\end{eqnarray}
whose numerical values are \cite[Table I]{CMNW-CGL35-Letter},

These two 
homoclinic patterns have the shape of a double well
(see \cite[Figures 2 and 3]{CMNW-CGL35-Letter})
%\ref{FigCGL5-Bound-state-1} and
%\ref{FigCGL5-Bound-state-2})
and they move with an arbitrary velocity if $p$ is real,
otherwise they are stationary.
They define two bound states of two CGL5 dark solitons,
as reported in \cite[Fig.~4]{ACM1998}.

%$\hbox{min}(M) $& $0.0908109634291956$&$2.23280070464833   $&$1.94651812146080  $&$0.0562997644388471$& \\ \hline 
%$M(\pm \infty) $& $.677170961403232  $&$2.44862453191137   $&$2.13466970112248  $&$.419823258180529  $& \\ \hline 
%$\hbox{max}(M) $& $1.03773444601239  $&$2.75846334543107   $&$2.40478194373398  $&$.643360405668782  $& \\ \hline 

The complex amplitude $A$ is the product of powers of six $\sinh$ functions
\begin{eqnarray}
& & 
\begin{array}{ll}
\displaystyle{
\frac{A}{K_0}=\hbox{rhs of  \cite[Eq.~(14)]{CMNW-CGL35-Letter}},
}
\end{array}
%\label{eqCGL5Trigo4a} 
\label{eqCGL5BoundA} 
\end{eqnarray}
in which $K_0$ is determined by the condition
$\lim_{\xi \to 0} \GLA \GLAc/M=1$.

%\vfill\eject
% ======================================================================
\subsection{CGL5 rational solution}

The subequation
\begin{eqnarray}
& & {\hskip -0.0 truemm}
%\left\lbrace
\begin{array}{ll}
\displaystyle{
\csi=e_r=d_r=g_i=0,g_r=\frac{3}{128} \left(1-5\sqrt{5} \right) \frac{d_i^2}{e_i}\ccomma
}\\ \displaystyle{
{M'}^4 - \frac{e_i^2}{3} \left[M + \left(1-\frac{5}{16} \left(1+\sqrt{5} \right) \right) \frac{d_i}{e_i}\right]^3 
                         \left[M +         \frac{3}{16} \left(1+\sqrt{5} \right)         \frac{d_i}{e_i}\right]^5=0,
%\ctwo=\frac{3}{16} \left(1+\sqrt{5} \right),
}
\end{array}
%\right.
%\label{eqCGL5Rat4}
\end{eqnarray}
has two branches (one for each sign of $\sqrt{5}$),
this is one of the binomial equations of Briot and Bouquet,
and its solution is rational
\begin{eqnarray}
& & {\hskip -0.0 truemm}
%\left\lbrace
%\begin{array}{ll}
%\displaystyle{
%(d_i \not=0):\
M=\frac{d_i}{e_i}\left[ - \frac{3(1+\sqrt{5})}{16} 
 - \frac{768 (2 + \sqrt{5})}{(d_i (\xi-\xi_0))^4/e_i^2-3 (8 (3+\sqrt{5}))^2}\right]\cdot
%}\\ \displaystyle{
%(\forall d_i):\
%\frac{M}{A_0^2}=\frac{1}{\xi} + \frac{i}{\xi-\xi_1} - \frac{1}{\xi-\xi_2} - \frac{i}{\xi-\xi_3} + c_0, A_0^8=\frac{3}{e_i^2},
%}\\ \displaystyle{
%(d_i=0):\ 
%\frac{M}{A_0^2}=\frac{1}{\xi-\xi_0}, d_i=0.
%\frac{e_i}{d_i}M= - \frac{3(1+\sqrt{5})}{16} 
% - \frac{768 (2 + \sqrt{5})}{(d_i \xi)^4/e_i^2-(8 \sqrt{3} (3+\sqrt{5}))^2}\cdot
%}
%\end{array}
%\right.
%\label{eqCGL5Rat4}
\end{eqnarray}
The limit $d_i=0, M=A_0^2/\xi$ is recovered after the translation
% xi0=2*sqrt(3)*(1+sqrt(5))/(di*A0^2), 
$\xi_0=2 \sqrt{3} (1+\sqrt{5})/(A_0^2 d_i)$.

The amplitude $A$ is the product of powers of six affine functions of $\xi$,
\begin{eqnarray}
& & {\hskip -0.0 truemm}
%\left\lbrace
\begin{array}{ll}
\displaystyle{
A=K_0 e^{\displaystyle -i \omega t + i \frac{\csr}{2} \xi} 
\left((d_i \xi)^2/e_i-8 i (1-\sqrt{5})\right)
}\\ \displaystyle{\phantom{123}
\left((d_i \xi)^2/e_i-8 \sqrt{3} (3+\sqrt{5})\right)^{(-1+i \sqrt{3})/2}
}\\ \displaystyle{\phantom{123}
\left((d_i \xi)^2/e_i+8 \sqrt{3} (3+\sqrt{5})\right)^{(-1-i \sqrt{3})/2},
K_0^2=- 3 d_i \frac{1+\sqrt{5}}{16 e_i}\cdot
%K0**2=-3*di*(1+sqrt(5))/16/ei
}
\end{array}
%\right.
%\label{eqCGL5Rat4}
\label{eqCGL5RatA}
\end{eqnarray}

The squared modulus is never bounded because two poles are real, therefore the solution is unphysical.

% ======================================================================
\section{Conclusion and perspectives}

In a previous version of the method \cite{ConteNgCGL5_ACAP},
for each solution we performed the independent integration of two elliptic equations,
namely the ODE for $M$ and the ODE for $(\log a)'$,
with the result that the two elliptic functions $\wp$ involved 
had different invariants $g_2,g_3$ linked by a Landen transformation.
The present method avoids this useless complication.

Searching for additional meromorphic traveling waves of CGL
is now proven to be hopeless.

The eleven meromorphic traveling waves of CGL require between one and five constraints
on the parameters of the ODE,
while the local representation of $M(\xi)$ by 
a Laurent series near one of its movable singularities does not require
any constraint.
In order to fill this gap,
i.e.~to build new closed form singlevalued traveling wave solutions,
necessarily nonmeromorphic, 
able to remove at least some of these constraints,
the guidelines
indicated by Painlev\'e in his ``Th\'eor\`eme g\'en\'eral''
\cite[pages 381--382]{PaiLecons}
and in his proof of the theorem of addition of Weierstrass
\cite[\S 41 p 51]{Pai-Thm-addition}
would certainly be quite useful.

This could provide an analytic description of the 
CGL3 homoclinic traveling hole,
observed in spatiotemporal intermittency \cite{vanHecke} 
but never found analytically.

At the PDE level, only one closed form solution is known: 
the CGL3 collision of two fronts \cite{NB1984},
\begin{eqnarray}
& & \hbox{(CGL3) }
\left\lbrace
\begin{array}{ll}
\displaystyle{
A=A_0 e^{-i \omega t}
   \left[\frac{k}{2}\sinh \frac{k}{2} x \right] \left[\cosh \frac{k}{2} x + e^{-(3/2) \gamma t}\right]^{-1+i \alpha},
}\\ \displaystyle{
c=0,
p_r=0,
k^2= - \frac{2 \gamma}{p_i},
\omega=- \frac{3 \gamma}{2},
}
\end{array}
\right.
\end{eqnarray}
which involves only one double pole of $M$.
It would be worth extending the present method to PDEs
and looking for solutions $M(x,t)$ with two poles (CGL3)
or at most four poles (CGL5).

% ============================================================================
\section*{Acknowledgements}

Constructive remarks of the referees greatly helped us to improve the manuscript.

The authors are pleased to thank Joceline Lega for sharing her expertise.
The first author thanks the Department of mathematics and the 
Institute of mathematical research of HKU,
and the Institute for Advanced Study of SZU
for support and hospitality.
The first two authors thank CIRM, Marseille (grant no.~2311) 
and IHES, Bures-sur-Yvette for their hospitality.
The third author was partially supported by the RGC grant no.~17307420. 
The first and last authors were partially supported by the 
National Natural Science Foundation of China (grant no.~11701382). 
The last author was partially supported by
Guangdong Basic and Applied Basic Research Foundation, China (grant no.~2021A1515010054). 

The authors have no competing interests to declare.

\vfill\eject



\textbf{List of all meromorphic solutions}
\medskip

The five solutions obtained by the present method are listed in section \ref{sectionNecessary}.
The six others are recalled in this Appendix for completeness.

% ------------------------------------------------------------------------
\appendix
% ======================================================================
\section{CGL3 source or propagating hole, pulse, front}

These three solutions 
(heteroclinic source or propagating hole \cite{BN1985} \cite[Fig.~5]{Lega2001},
 homoclinic pulse \cite{HS1972}, and
 heteroclinic front \cite{NB1984})
possess a subequation with only one double pole,
for the respective parameter values,
\begin{eqnarray}
& & {\hskip -25.0 truemm}
\hbox{(source/hole) }
%\left\lbrace
%\begin{array}{ll}
%\displaystyle{
g_i=\frac{2}{3 \alpha} g_r-\frac{1+\alpha^2}{9 \alpha^2} \csi^2,
%}\\ \displaystyle{
%\left\lbrack M'-\frac{2}{3} \csi \left(M-\frac{\csi^2}{3 d_i \alpha}\right)\right\rbrack^2 
%       -\frac{4 d_i}{3 \alpha}   \left(M-\frac{\csi^2}{3 d_i \alpha}\right)
%                                 \left(M+\frac{\csi^2}{3 d_i \alpha} - \frac{g_r}{d_i}\right)^2=0,
%}
%\end{array}
%\right.
\label{eqCGL3Hole}
\end{eqnarray}
\begin{eqnarray}
& & {\hskip -35.0 truemm} \hbox{(pulse) }
\csi=0,\
g_i=\frac{1-\alpha^2}{2 \alpha} g_r,\
%{M'}^2 -\frac{4 d_i}{3 \alpha} \left(M- \frac{3 g_r}{2 d_i} \right) M^2=0,
\label{eqCGL3Pulse}
\end{eqnarray}
\begin{eqnarray}
& & {\hskip -45.0 truemm} \hbox{(front) }
g_r=0, g_i=\frac{2}{9}\csi^2.
%\left\lbrack M'-\frac{2}{3} \csi M \right\rbrack^2 -\frac{4 d_i}{3 \alpha} M^3=0.
\label{eqCGL3Front}
\end{eqnarray}
\medskip

The homoclinic or heteroclinic physical bounded expressions
for the complex amplitudes are,
\begin{eqnarray}
& & \frac{\GLA}{K_0} =
e^{\displaystyle{i[\alpha \log \cosh \frac{k}{2} \xi -\omega t +\frac{\csr}{2} \xi]}}
\times                         
\left\lbrace
\begin{array}{ll}
\displaystyle{
\hbox{(source) }
\left[\frac{k}{2} \tanh \frac{k}{2} \xi +(X+i Y) c \right] e^{\displaystyle{i K c \xi}},\
\label{eqCGL3HoleA} % = source
}\\ \displaystyle{
\hbox{(pulse) }
(-i k \sech k x), c=0,
\label{eqCGL3PulseA}
}\\ \displaystyle{
\hbox{(front) }
\frac{k}{2} \left[\tanh \frac{k}{2} \xi \pm 1 \right] e^{\displaystyle{i K c \xi}},
\label{eqCGL3FrontA}
}
\end{array}
\right.
\end{eqnarray}
in which the real constants $X,Y,K$ only depend on $p,q,\gamma$,
see for instance \cite{CM1993}.

When applied to a suitable complex variable \cite{CM1993},
the truncation procedure \cite{WTC} generates one short complex relation
between the three real parameters $\omega,c^2,k^2$,
\begin{eqnarray}
& & \frac{i \gamma- \omega}{p}= \left(\frac{c}{2 p}\right)^2 +
\left\lbrace\begin{array}{ll}
    \displaystyle{\hbox{(source) } (3 i \alpha-2) \frac{k^2}{4}\ccomma
}\\ \displaystyle{\hbox{(pulse) } (1-i \alpha)^2 k^2,
}\\ \displaystyle{\hbox{(front) } \frac{k^2}{4}\cdot
}\end{array}\right.
\end{eqnarray}

%\vfill\eject

% ======================================================================
\section{CGL5 front, source or sink, pulse}

These three solutions 
(a heteroclinic front \cite{vSH1992},
 a homoclinic source/sink \cite{Moores,MCC1994},
 a homoclinic pulse \cite{vSH1992})
share a common simplicity when they are described 
by the truncation \cite{WTC} of the 
suitable complex variable already mentioned \cite{CM1993}.

Under the following constraints and values of $k^2, \taub$,
\begin{eqnarray}
& & {\hskip -12.0 truemm} \hbox{(front) }
\left\lbrace
\begin{array}{ll}
\displaystyle{
%ogrvalbis=-(csi+2*a2*dr+4*tau1)*(csi+2*tau1)/(4*alpha);
%ogivalbis= (csi+2*a2*dr+4*(1-2*alpha**2)*tau1)**2/(4*alpha)**2
%      +tau1*(2*a2*dr+(3-4*alpha**2)*tau1);		
%tau1val=-csi/4+a2*(2*alpha*di-dr)/(2*den1);
%den1val= 1+a2**4*ei**2;
g_r=-\frac{(\csi+2 A_0^2 d_r+4 \tauone)(\csi+2\tauone)}{4 \alpha},
\tauone=-\frac{\csi}{4}+A_0^2 \frac{2 \alpha d_i-d_r}{2(1+4 \alpha^2)},
k^2= 4 \tau_1^2,
}\\ \displaystyle{
g_i= \frac{(\csi+2 A_0^2 d_r+4 (1-2\alpha^2)\tauone)^2}{(4 \alpha)^2}
    +\tauone (2 A_0^2 d_r+(3-4\alpha^2)\tauone),
%}\\ \displaystyle{
%{M'} + A_0^{-2} M (M + 2 A_0^2 \tauone)=0,\
}
\end{array}
\right.
\label{eqCGL5Front}
\end{eqnarray}
\begin{eqnarray}
& & 
\hbox{(source) }
\left\lbrace
\begin{array}{ll}
\displaystyle{
\csi=0,
 \frac{g_i}{(3+2 \alpha^2) d_i + 5 \alpha d_r}
=\frac{g_r}{12 \alpha (d_i + 2 \alpha d_r)}
=\frac{A_0^4 [(1-2 \alpha^2) d_i + 3 \alpha d_r]}{4 \alpha^2 (1+4\alpha^2)^2 },\
%}\\ \displaystyle{
%g_i=  M_0 \frac{(3+2 \alpha^2) d_i + 5 \alpha d_r}{4 \alpha D_1},\
%g_r=3 M_0 \frac{               d_i + 2 \alpha d_r}{         D_1},\
%M_0=A_0^4 \frac{(1-2 \alpha^2) d_i + 3 \alpha d_r}{  \alpha D_1},\
%D_1=1+4 \alpha^2,\
%}\\ \displaystyle{
%{M'}^2 - A_0^{-4} M
%\left\lbrack M + 2 A_0^2 \tau_d\right\rbrack^2
%\left\lbrack M + 2 A_0^2 (\tau_b+\tau_d) \right\rbrack
%=0,\
}\\ \displaystyle{
k^2=2 \frac{g_r}{\alpha} - 4 g_i,
\coth\frac{k \xib}{2}=\tau_b= A_0^2 \frac{(3-2\alpha^2)  d_i + 7 \alpha d_r}{2 \alpha (1+4 \alpha^2)},\
%\tau_d= A_0^2 \frac{(2 \alpha^2-1) d_i - 3 \alpha d_r}{2 \alpha D_1},\
}
\end{array}
\right.
\label{eqCGL5Source}
\end{eqnarray}
\begin{eqnarray}
& & {\hskip -30.0 truemm} \hbox{(pulse) }
\left\lbrace
\begin{array}{ll}
\displaystyle{
\csi=0,\
d_r=\frac{(2 \alpha^2-1)}{3 \alpha} d_i,\
g_i=\frac{(1-4 \alpha^2)}{4 \alpha} g_r, % \not=0,\
}\\ \displaystyle{
%{M'}^2 - A_0^{-4} M^2 \left\lbrack M^2 + \frac{2 d_i A_0^4}{3 \alpha} M-\frac{A_0^4}{\alpha} g_r \right\rbrack=0,\
k^2=-\frac{g_r}{\alpha},
\coth\frac{k \xib}{2}=\taub,
k\left(\frac{k}{2 \tau_b}+\frac{2 \tau_b}{k}\right)=\frac{2 A_0^2 d_i}{3 \alpha},
}
\end{array}
\right.
\label{eqCGL5Pulse}
\end{eqnarray}
the bounded, physical expressions
(heteroclinic front, homoclinic source and pulse)
of the complex amplitudes are,
%result from the choice $\xi_0=(2/k) i \pi/2$,
\begin{eqnarray}
& & {\hskip -12.0 truemm} \frac{\GLA}{A_0} =
e^{\displaystyle{i\left[-\omega t +\frac{\csr}{2} \xi\right]}}
\times                         
\left\lbrace
\begin{array}{ll}
\displaystyle{
\hbox{(front) }
\left(\frac{k}{2} \left(\tanh \frac{k \xi}{2} + 1\right) \right)^{1/2}
e^{\displaystyle{i\left[\alpha \log \cosh \frac{k \xi}{2} + \Kshift \xi\right] }},
\label{eqCGL5FrontA}
}\\ \displaystyle{
\hbox{(source) }
\left(\frac{k \sinh k b}{\cosh k \xi - \cosh k b} + k \tanh \frac{k b}{2}\right)^{1/2}
e^{\displaystyle{i[\alpha \log(\cosh k \xi - \cosh k b)]}},
\label{eqCGL5SourceA} % = source
}\\ \displaystyle{
\hbox{(pulse) }
\left(\frac{k \sinh(k b)}{\cosh k \xi - \cosh k b} \right)^{1/2}
e^{\displaystyle{i[\alpha \log(\cosh k \xi - \cosh k b)]}}. 
\label{eqCGL5PulseA}
}
\end{array}
\right.
\end{eqnarray}

The most compact expressions characterizing the parameters of these solutions
are obtained in complex notation \cite{CM2005},
these are:
the definition of $\alpha$ and $A_0^2$,
\begin{eqnarray}
& & r=-\frac{p}{A_0^4} \left(-\frac{1}{2}+i \alpha\right) \left(-\frac{3}{2}+i \alpha\right),
\end{eqnarray}
and two additional relations,
\begin{eqnarray}
& &    {\hskip -23.0 truemm}              
%\left\lbrace
\begin{array}{ll}
\displaystyle{
\hbox{(front) }
q=\frac{2 i p}{A_0^2}\left(-\frac{1}{2}+i \alpha\right) 
 \left[\Kshift+\frac{\csr}{2}-\frac{c}{2 p}+\frac{k}{4}(2 \alpha+3 i)\right],
%}\\ \displaystyle{
\frac{i \gamma- \omega}{p}=
\left(\frac{c}{2 p}\right)^2
-\left(\Kshift+\frac{\csr}{2}-\frac{c}{2 p}+\frac{k}{4}(2 \alpha+ i) \right)^2,
%\label{eqCGL5Front}
}\\ \displaystyle{
\hbox{(source) }
q=\frac{2 k p}{A_0^2 \sinh k b} \left(-\frac{1}{2}+i \alpha\right) 
\left(2 - i \alpha - \cosh k b\right),
%}\\ \displaystyle{
%write 0==let i**2=-1
% in (c/(2*p))**2+(k/2)**2+(-1/2+i*alpha)*(3/2)*(k/cosh(k*b/2))**2
%    -(i*gammafinal-omega)/p;		
\frac{i \gamma- \omega}{p}=
\left(\frac{c}{2 p}\right)^2
+\left(\frac{k}{2}\right)^2
+\frac{3}{2}\left(-\frac{1}{2}+i \alpha\right) \left(\frac{k}{\cosh \displaystyle\frac{k b}{2}}\right)^2,
%\label{eqCGL5Source} % = source
}\\ \displaystyle{
\hbox{(pulse) }
%\hbox{Below, new physical formulae $c=0$. Old = MCC 1994 (24), CM 2005 (34) $r_0=0$}
q=-\frac{p}{A_0^2}\left(-\frac{1}{2} + i \alpha \right)	(-1+i \alpha) \ 2 k \coth(k b),
\frac{i \gamma-\omega}{p}=\left(\frac{c}{2 p}\right)^2
 +\left(-\frac{1}{2} + i \alpha \right)^2 k^2.
%K=\frac{c}{2 p}.
%\label{eqCGL5Pulse}
}
\end{array}
%\right.
\end{eqnarray}

\vfill\eject



\vfill\eject
\end{document}